\documentclass[12pt]{article}   

\usepackage{graphicx}
\usepackage{scicite}

\usepackage{times}

\newenvironment{sciabstract}{%
\begin{quote} \bf}
{\end{quote}}

\newcounter{lastnote}
\newenvironment{scilastnote}{%
\setcounter{lastnote}{\value{enumiv}}%
\addtocounter{lastnote}{+1}%
\begin{list}%
{\arabic{lastnote}.}
{\setlength{\leftmargin}{.22in}}
{\setlength{\labelsep}{.5em}}}
{\end{list}}

\title{Heteronuclear Cooper Pairs in An Ultracold Atomic Gas}

\author
{Yongle Yu,$^{\ast}$\\
\\
\normalsize{Mathematical Physics, Lund Institute of Technology,
 SE 221 00 LUND, Sweden}\\
\normalsize{$^\ast$To whom correspondence should be addressed;
 E-mail:  yongle.yu@matfys.lth.se.}
}

\date{}

\begin{document} 


\baselineskip24pt


\maketitle


\begin{sciabstract}
In an ultracold mixture of two different Fermi species of
atoms,  Cooper pairs can be formed between two
different atoms. The masses of one atom and
its partner in this kind of Cooper pairs may differ by order
of magnitude.
In this system,  each species of atoms are in the same
 atomic spin state and two species have the same atomic densities.
The pairing gap diminishes if two species have different densities
and vanishes when the density imbalance reaches a critical value.
\end{sciabstract}



Ultracold atomic gases offer an unprecedented opportunity to study 
many body physics with short range interaction in the sense that
atomic species, density, interaction strength and trapping 
potential can be altered.  Following the achievement of
Bose-Einstein Condensation in a Bose gas of atoms, 
experimenters made rapid progress in manipulating an ultracold degenerate
Fermi gas of atoms \cite{choSCI} and
observed a superfluid and condensate phase 
 \cite{jochimSCI,greinerNA,zwierleinPRL1,RegalNA,bourdelPRL,
huletPRL, chinSSC} when fermionic atoms interact near Feshbach resonance.
Below Feshbach resonance ($a < 0$, $a$ is the $s$-wave scattering length),
 Cooper pairs
are formed due to the many body effect while the 
two body interaction is strong
enough to generate weakly bound molecules (dimers) above Feshbach resonance
($a > 0$).   
Being of bosonic nature, Cooper pairs
and dimers
alter the quantum statistics of the system and
modified behaviours of the system fundamentally
near zero temperature.  By tuning the interaction strength,
experiments  \cite{bartensteinPRL1, regalPRL,zwierleinPRL2,
kinastPRL,bartensteinPRL2, chinSSC}  
  began to explore BCS-BEC crossover, which is of
great theoretical interest for decades \cite{eaglesPR, 
leggett,nozieres, melo}. Some theoretical treatments of
pairing and BCS-BEC crossover in a Fermi gas of atoms can
be found in \cite{holland, timmermans, heiselberg, ohashi,
milstein, combescot, bulgac, carlson, ho}.

In all systems, Cooper pairs which have been studied
 till now are symmetrical pairs in the sense
that two partners in a pair are of the same species and may 
differ only by the projection of
spin or pseudospin. In nuclear physics, Cooper pairs
composed of a neutron and a proton is considered.
 However, neutrons and
protons can be considered as one species, the nucleons
 due to well known isospin invariance of nuclear 
interaction.  In this paper, we point out
 Cooper pairs composed 
of two different species of atoms can occur in 
an ultarcold gas. The masses of two
partners in such unconventional Cooper pairs
may differ by order of magnitude 
(for example, Cooper pairs
formed by  ${^6}Li$ atoms and  ${^{40}}K$ atoms).

We consider a mixture of two different species of fermionic atoms, 
denote one species by $A$ , the other by $B$ and their atomic masses
by $m_A$ and $m_B$, respectively. 
All atoms of species $A$ are set to be 
in the same atomic hyperfine state,
for example, the lowest 
Zeeman spin state in a presence of magnetic field, so are the 
atoms of species $B$.  Preparing one species of 
polarized atoms is not a problem, see 
\cite{dieckmann} for example.    
the intraspecific interaction in the 
mixture can be neglected
for the following reason. The average 
interatomic distance is orders of magnitude
 larger than the 
interaction 
range in this dilute system and
the interaction is well approximated by a zero-range
model. Pauli principle prevents two atoms of the same
species and of the same spin state from
approaching each other and therefore they hardly
interact.
Mathematically, it is easy to show that  
the matrix elements of a zero-range interaction vanish in
a basis of many body states with polarized spins,
$ \langle \psi_{j1}({\bf r}_1,{\bf r}_2, ...)\mid g 
\delta({\bf r}_1-{\bf r}_2) \mid \psi_{j2}({\bf r}_1,{\bf r}_2, ...)
\rangle = 0$ ( $g$ is
the interaction strength). The wavefunctions are
 antisymmetric under exchange
of ${\bf r}_1$ and ${\bf r}_2$,  which leads to 
$ \psi_{j1}({\bf r}_1,{\bf r}_1, ...)=-\psi_{j1}
({\bf r}_1,{\bf r}_1, ...)
= 0 $ and $  \psi_{j2}({\bf r}_1,{\bf r}_1, ...)=
-\psi_{j2}({\bf r}_1,{\bf r}_1, ...) = 0$ . 

 The dominating interaction in the system is the interspecific 
scattering $ V_{AB}({\bf r})= \frac { 2 \pi\hbar^2 a_{AB}}
{m_{AB}} \delta ({\bf r})$, where $ a_{AB}$ is the s-wave scattering 
length between two species of atoms and $ m_{AB}$ is 
the reduced mass, $ m_{AB}= \frac {m_A m_B}{m_A + m_B}$
. A 
weak attractive interspecific scattering ($a_{AB} < 0$) can cause Cooper
instability of normal phase and lower the energy of the system
by forming pair correlations between two species of atoms. Note
the interatomic force doesn't influence the atomic spins at 
this low energy regime. To illustrate condition for Cooper
 pairing and to study
the equation for the pairing gap, let us consider
a BCS wave function  of the system. For simplicity we only consider
a homogeneous system,
\begin{equation}
 \mid  \psi_{BCS} \rangle = \prod_{\bf k} ( u_{\bf k} + 
 v_{\bf k}a_{\bf k}^\dagger b_{ \bar{\bf k}}^\dagger ) 
\mid vac. \rangle
\label{eq:bcs}
\end{equation}
Where $a^\dagger$ and $b^\dagger$ are creation operators for
 atoms of species A
and atoms of species B, respectively.
 Normalization of this wavefunction
requires,
\begin{equation}
\mid u_{\bf k}\mid ^2 +  \mid v_{\bf k}\mid ^2 = 1
\label{eq:norm}
\end{equation}
$ \mid v_{\bf k} \mid ^2$ is the occupation 
probability of state ${\bf k} $ by the atoms
of species A.

Cooper pairs have zero momentum, which implies $ \hbar {\bf k}
+  \hbar \bar{\bf k}= 0 $ and $ \bar{\bf k}= -  {\bf k}$. 
 Two species have the same particle density distribution
in the momentum space ($\mid v_{\bf k}\mid ^2 =\mid v_{-\bf k}\mid ^2$,
as we will see it later)   and thus the same densities ($n_A= n_B $).
This density relationship is a condition for Cooper pairing and 
enables a complete pairing correlation between
 two species of atoms.  It is remarkable
that two species share the same Fermi momentum rather than the same
Fermi energy, which can be easily derived from 
the density relationship.

A 'ground state' in normal phase corresponds to a sharp 
particle distribution
near the Fermi surface,
\[ \left\{ \begin{array}{ll}
 \mid u_{\bf k}\mid ^2 = 1 & k < k_F\\
 \mid u_{\bf k}\mid ^2 = 0 & k > k_F\\
 \end{array} \right. \]
With $k_F$ is the Fermi wave number,$k_F=  (6 \pi^2 n_A )^{1/3}$.  
A superfluid ground state with  a smooth particle distribution
near the Fermi surface can further lower the energy of the system.

The equation for the gap at zero temperature 
can be derived using variational principle, similar to what
is done in the case of a system with symmetrical Cooper pairs
\cite{tinkham}. The
Hamiltonian with pairing interaction is written as,
\begin{equation}
 H = \sum_{\bf k}  ( \epsilon_{\bf k}^A a_{\bf k}^\dagger a_{\bf k}  + 
\epsilon_{\bf k}^B b_{\bf k}^\dagger b_{\bf k}) +
\sum_{\bf k l}
V_{{\bf k}{\bf l}}  a_{\bf k}^\dagger b_{- \bf k}^\dagger
 a_{\bf l} b_{- \bf l}
\label{eq:hami} 
\end{equation}
Where $\epsilon_{\bf k}^A$ and $\epsilon_{\bf k}^B$ are the single
particle energies in the normal phase,  $\epsilon_{\bf k}^A = 
\frac{ \hbar^2 k^2 }{2 m_A} $ and $\epsilon_{\bf k}^B = 
\frac{ \hbar^2 k^2 }{2 m_B} $ ( precisely, there is 
a mean field shift in $\epsilon_{\bf k}^A$ and $\epsilon_{\bf k}^B$.
The shift can be neglected in discussion of the equation for the gap).

Since a BCS state is not an eigenstate of particle number operators,
restrictions on the average number of particles are imposed,
\begin{equation}
\bar N_A \equiv \langle   \psi_{BCS} 
  \mid  \sum_{\bf k}
 a_{\bf k}^\dagger a_{ \bf k}\mid \psi_{BCS} \rangle = N
\label{eq:chemi1}
\end{equation}
\begin{equation}
\bar N_B \equiv \langle   \psi_{BCS} 
  \mid  \sum_{\bf k}
b_{\bf k}^\dagger b_{ \bf k} \mid \psi_{BCS} \rangle = N
\label{eq:chemi2}
\end{equation}
where $N$ is the number of atoms of each species ($N= N_A = N_B$).

To minimize the ground state 
energy $ E\equiv  \langle  \psi_{BCS} \mid
 H \mid \psi_{BCS} \rangle$ 
subject to Eqs.(\ref{eq:chemi1}, \ref{eq:chemi2}), 
 chemical potentials
 are introduced as Lagrange parameters in finding 
the absolute minimum of
\begin{equation}
W \equiv
E - \mu_A \bar N_A -  \mu_B \bar N_B 
= 
 \langle   \psi_{BCS} 
  \mid H - \mu_A \sum_{\bf k} a_{\bf k}^\dagger a_{ \bf k}
 -\mu_B \sum_{\bf k} b_{\bf k}^\dagger b_{ \bf k} \mid \psi_{BCS} \rangle
\label{eq:forw}
\end{equation}
where $\mu_A$ and  $\mu_B$ are chemical potentials for
atoms of species $A$ and species B, respectively.

Combining Eqs.(\ref{eq:bcs}, \ref{eq:norm}, \ref{eq:hami})and (\ref{eq:forw})
  , and taking $u_{ \bf k}$,
 $v_{ \bf k}$ to be real and $u_{ \bf k}$ 
non negative for simplicity, 
we have,

\begin{equation}
W = 2\xi_{\bf k} \sum_{\bf k}
v_{\bf k}^2 +  \sum_{{\bf k}{\bf l}} V_{{\bf k}{\bf l}}
\sqrt{1 - v_{\bf k}^2}v_{\bf k}\sqrt{1 - v_{\bf l}^2} v_{\bf l}
\end{equation}
with 
$\xi_{\bf k}= \frac{1}{2} (\epsilon_{\bf k}^A + \epsilon_{\bf k}^B
- \mu_A - \mu_B) = \frac{\hbar^2 ( k^2 - k_F^2)}{4 m_{AB}}$ (
$\mu_A \approx \frac{\hbar^2 k_F^2}{2 m_{A}}$, 
$\mu_B \approx \frac{\hbar^2 k_F^2}{2 m_{B}}$). 
   
 $ \frac{\delta W}{\delta v_{\bf k}}= 0$ ,  which lead to the following
 equation for the gap,

\begin{equation}
\Delta_{\bf k}= - \frac{1}{2} \sum_{\bf l} V_{{\bf kl}}  \frac 
{\Delta_{\bf l}}{E_{\bf l}}
\label{eq:gap}
\end{equation}
Where $\Delta_{\bf k} $ is gap parameter,
\begin{equation}
\Delta_{\bf k}= - \sum_{{\bf l}}  V_{\bf kl} u_{\bf l} v_{\bf l} 
\end{equation}
and 
\begin{equation}
E_k= \sqrt {\xi_{\bf k}^2    + \Delta_{\bf k}^2  }
\end{equation}

In terms of
 $\Delta_k $ and $E_k$, the fractional occupation number $v_k^2 $ is given by,
\begin{equation}
v_{\bf k}^2= \frac {1}{2}
(1 -  \frac{\xi_{\bf k}}{ E_{\bf k}} ) 
\end{equation}

As the Fourier transfer of $ V_{AB}({\bf r})$, the pairing 
interaction is,
\begin{equation}
V_{\bf kl}=   \frac { 2 \pi\hbar^2 a_{AB}} {m_{AB}}
\label{eq:fourier}
\end{equation}
With Eqs. (\ref{eq:gap}) and (\ref{eq:fourier}),
  Gap parameter then is a constant function of ${\bf k}$
($\Delta_{\bf k}= \Delta$.  Physically, for
 a short range 
interaction, $V_{\bf kl}$ is approximately a constant function
of ${\bf k}$ and ${\bf l} $ when ${ \hbar\bf k},{ \hbar\bf l} $ lie
in the pairing space near Fermi surface). Given $V_{\bf kl}$
is a constant for any ${\bf k}$ and ${\bf l}$, however,
 the sum over the 
spectrum
in the right side of Eq.(\ref{eq:gap}) formally diverges.
This issue can be handled either by imposing 
a cutoff on the spectrum or  by taking
a renormalization approach.
Since there is no physical reason for introducing an energy cutoff
for this system, we take a natural regularization 
scheme \cite{bulgac2, bruun} to remove the divergence.  Using a formula
derived in \cite{bulgac2} with some replacements,
we have the following equation for the gap,
\begin{eqnarray}
& & \frac{1}{k_F}\int _0^{k_c}dk
\frac{\hbar^2 k
^2}{\sqrt{\hbar^4 (k^2-k_F^2)^2+ 4 m^2 \Delta ^2 }}
\nonumber  \\
& & =
\frac{\pi}{2k_F|a_{AB}|}
 \left [
  1 +\frac{2k_c|a_{AB}|}{\pi} -\frac{k_F|a_{AB}|}{\pi}\ln\frac{k_c+k_F}{k_c-k_F}
 \right ]
 \label{eq:gap2} ,
\end{eqnarray}

\begin{figure}
\begin{center}\
\includegraphics[bb= 105 299 475 496]{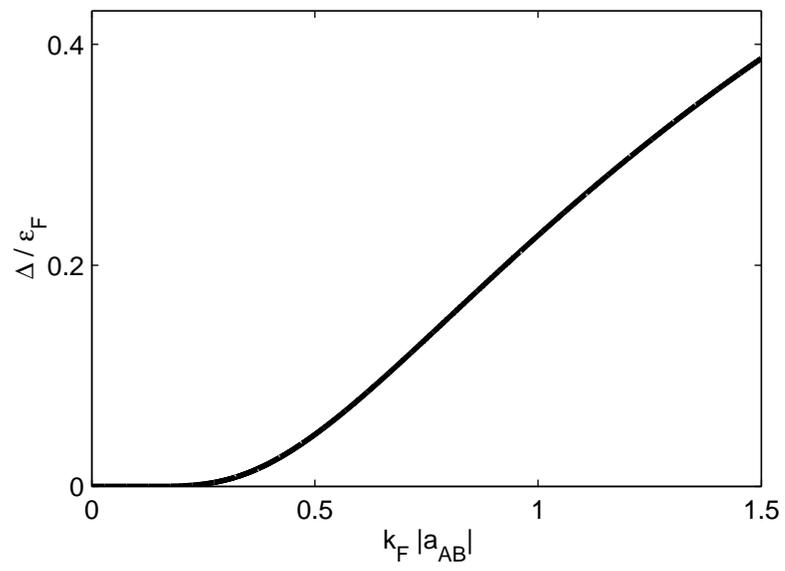}
\end{center}
\caption{The pairing gap as a function of $k_F\mid a_{AB}\mid$, 
 $\epsilon_F= 
\frac { \hbar^2 k_F^2 }{4 m_{AB}}$. 
\label{fig:gap}}
\end{figure}

Where $k_c $ is the cutoff on wave number. $\Delta$ is essentially
 independent of  $k_c $ when  $k_c $ is big enough. We numerically
computed $\Delta $ as a function of  $k_F |a_{AB}| $ and plotted it
in the  Fig.(~\ref{fig:gap}).

In atomic gas, the interaction can be tuned. Observations
of heteronuclear Feshbach resonance in a mixture of bosons
and fermions  were reported 
recently in \cite{stanPRL, inouyePRL}.
For the system we considered above, The phenomenon of
BCS-BEC crossover can  be explored if Feshbach resonance 
between two species is found.  Dimers composed of an
 one atom of species $A$ and one  atom of species $B$ 
are formed in the BEC regime.
   
In the BCS regime, a naturally raised question is 
what is the case if $n_A \neq n_B $?
 Take for example $N_A > N_B$, some atoms of species
$A$ are not paired. The unpaired atoms  occupy some states 
near the Fermi surface and the availability of the states for pair
scattering is reduced, consequently the pairing gap diminishes. As a 
monotonic decreasing function
of  density imbalance  $ \chi= \mid\frac{n_A - n_B}{n_a + n_B}\mid $, 
the pairing
gap vanishes when $ \chi $ reaches a critical value. Such a density
imbalance induced phase transition
(for at least one species of atoms) can be studied in an atomic gas and
shall enhance our understanding of pairing correlations.

\bibliography{scibib}

\bibliographystyle{Science}


\begin{scilastnote}
\item {\bf Achowledgements} This work was supported by 
the Swedish Foundation for
strategic Research.
\end{scilastnote}


\clearpage

\end{document}